\documentclass[journal,twoside,11pt]{IEEEtran}
\usepackage[utf8]{inputenc}
\usepackage[T1]{fontenc}
\usepackage{graphicx}
\usepackage{amsmath,amssymb,amsfonts}
\usepackage{textcomp}
\usepackage{siunitx}
\usepackage{booktabs}
\usepackage{cite}
\usepackage{url}
\usepackage{xcolor}
\usepackage{subcaption}
\usepackage{tabularx}
\usepackage{array}
\usepackage[hidelinks]{hyperref}

\newif\ifrevcolor
\revcolorfalse % <<< final (black) version
\ifrevcolor
 \newcommand{\rev}[1]{\textcolor{red}{#1}}
\else
 \newcommand{\rev}[1]{#1}
\fi

\begin{document}

\title{Circuit-Level Design, Modeling, and On-Wafer Characterization of a Coplanar THz Optoelectronic Mixer}

\author{S.~Islam, V.~Merupo, C.~Chamek, M.~Sassi, C.~Coinon, Y.~Deblock, E.~Okada,
 S.~Lepilliet, F.~Braud, Q.~Fornasiero, G.~Di~Gioia, M.~Faucher, G.~Ducournau, G.~Loas, S.~Arscott, and~E.~Peytavit

 \thanks{Manuscript received \today.
This work was supported by the France 2030 PEPR ``Electronics''
program through the FUNTERA project (ANR-22-PEEL-0006); by the
Contrat de Plan État--Région (CPER) WaveTech; by the C2EMPI project
(R-CDP-24-004-C2EMPI); by the Start-AIRR ASPIR project of the
Hauts-de-France Regional Council; by the French National Research
Agency (ANR) through the PISA project (ANR-23-CE42-0025); by the
RENATECH+ national microfabrication network; and by the RF-Net network (ANR- 22-PEFT-0011).}

 \thanks{S.~Islam, V.~Merupo, C.~Chamek, M.~Sassi, C.~Coinon, Y.~Deblock, E.~Okada,
 S.~Lepilliet, F.~Braud, Q.~Fornasiero, G.~Di~Gioia, M.~Faucher, G.~Ducournau, S.~Arscott, and~E.~Peytavit are with IEMN, Univ.\ Lille, CNRS, Univ.\ Polytechnique
 Hauts-de-France, UMR~8520, F-59000 Lille, France
 (e-mail: \texttt{emilien.peytavit@univ-lille.fr}).}
 \thanks{G.~Loas is with Institut Foton, UMR 6082 / CNRS - Univ Rennes – INSA, France.}}

\markboth{IEEE Transactions on Terahertz Science and Technology,~Vol.~XX, No.~X, Month~20XX}%
{Islam \MakeLowercase{\textit{et al.}}: Circuit-Level Design, Modeling, and On-Wafer Characterization of a Coplanar THz Optoelectronic Mixer}

\maketitle

% =========================================================================
\begin{abstract}
We report the design, modeling, and on-wafer characterization of a
photoconductive heterodyne mixer implemented as a true
terahertz monolithic integrated circuit (TMIC). Unlike previous
photoconductive heterodyne demonstrations based on antenna coupling or
direct probe measurements without controlled RF/IF embedding, the
proposed circuit is characterized using the figures of merit normally
expected from an RF integrated mixer, including input return loss,
broadband impedance matching, RF/IF port isolation, and quantitatively
optimized conversion loss. We
design a fully coplanar mixer combining a microcavity photoconductor
with a broadband matching network (Ti shunt resistor), an RF DC-block
(series MIM capacitor) and an RF/IF decoupling element (shunt MIM
capacitor), and we develop a parametric circuit-level model that
captures both the time-varying photoconductance and the measured CPW
embedding. The passive part of the model is validated up to
\SI{500}{\giga\hertz} by VNA measurements of the unilluminated
($R_\mathrm{off}\!\sim\!\SI{1}{\mega\Omega}$) full TMIC, showing good agreement with the circuit simulation using
the same off-state device. {This anchors the predictive value of the simulation under
operating conditions: simulated with the lowest on-state resistance
measured at the maximum applied optical pump power
($R_\mathrm{on}\!\sim\!110\,\Omega$), $|S_{11}|$ remains below
$-10$\,dB up to \SI{215}{\giga\hertz} and below
$-6.5$\ up to \SI{500}{\giga\hertz}, which, although not optimal,
constitutes an acceptable matching level at these frequencies.}
Measured and simulated conversion
losses agree within $\pm \SI{1}{\decibel}$ from \SI{1}{\giga\hertz} up
to \SI{500}{\giga\hertz}, with \rev{$\tau \approx \SI{0.8}{\pico\second}$
 and the computed
$C_\mathrm{pc} \approx \SI{10.8}{\femto\farad}$}. The conversion loss
is \SIrange{22.5}{25.5}{\decibel} below \SI{320}{\giga\hertz} and
\SIrange{26}{34}{\decibel} up to \SI{500}{\giga\hertz}. The TMIC is fabricated on a \SI{10}{\micro\meter} SOI device layer,
preserving compatibility with future backside processing for
membrane-supported modules incorporating rectangular-waveguide
transitions or heterogeneous integration onto micromachined silicon
probes.
\end{abstract}


% =========================================================================
\section{Introduction}
\IEEEPARstart{C}ontinuous-wave (CW) terahertz detection using ultrafast photoconductors has been a cornerstone of room-temperature THz spectroscopy since the pioneering work of Brown, McIntosh, and Verghese~\cite{Verghese1998}. In CW photomixing systems, two single-mode lasers detuned by a terahertz-frequency offset generate an optical beat note that simultaneously pumps two antenna-coupled photoconductive devices. One device is electrically biased to generate THz radiation (the photomixer), while the second operates unbiased to enable coherent detection, together forming the core of a THz spectrometer. CW photoconductive systems now routinely deliver dynamic ranges above \SI{100}{\decibel} and usable bandwidths up to several THz. At telecom wavelengths around \SI{1550}{\nano\meter}, uni-traveling-carrier photodiodes (UTC-PDs) have come to dominate the emission side due to their superior output power and conversion efficiency, while photoconductors remain the standard detection element~\cite{Roggenbuck2015,Deninger2015,Kohlhaas2019}. The same photoconductive principle naturally extends to heterodyne mixing, in which an external RF/THz signal is downconverted to a low intermediate frequency read out by standard electronics; this has been demonstrated mainly at 780-nm and 1550-nm wavelengths using low-temperature-grown GaAs and Iron-doped-InGaAs materials, respectively~\cite{Peytavit2013,Tannoury2023,DeumerAPL2025,Deumer2025} . 

In every reported photoconductive heterodyne demonstration,
the conversion loss reported over the 100--500\,GHz band
lies between 20 and 35\,dB~\cite
{Peytavit2013,Tannoury2023,DeumerAPL2025,Deumer2025}.
This remains larger than the effective single-sideband conversion
losses of commercially available Schottky fundamental and
sub-harmonic mixers, whose specified double-sideband conversion losses
are typically in the \SIrange{8}{11}{\decibel} range over
\SIrange{220}{500}{\giga\hertz}~\cite{VDI_mixers}. After conversion to
the present single-tone definition, this corresponds to approximately
\SIrange{11}{14}{\decibel}. However, such electronic mixers generally
require high-frequency LO chains, narrow-band waveguide circuits, and
careful RF/LO filtering.

Commercially available even-harmonic mixers (EHMs) partially relax the
LO-frequency constraint by operating with relatively low-frequency
electronic local oscillators and high harmonic orders (typically
16--20 in the 220--500\,GHz range). However, their conversion loss
remains high, typically \SIrange{30}{45}{\decibel} in commercial
modules~\cite{VDI_mixers}, comparable to or larger than that of the
present photoconductive mixer . Such high-order harmonic mixing generates multiple spurious
conversion products, making frequency planning and filtering more
challenging at the system level.

Photonic heterodyne approaches based on UTC photodiodes naturally
avoid electronic THz LO generation by directly producing the LO signal
from an optical beat note distributed through optical
fiber~\cite{Rouvalis2012,BelioApaolaza2022}. In such approaches, the
photodiode either operates directly as an optoelectronic mixer through
its nonlinear photocurrent response, or generates the RF/THz LO power
used to pump a Schottky mixer. In both cases, efficient operation requires large photocurrents
generated by the optical beat note. These photocurrents increase shot
noise and transfer optical pump noise to the RF domain. 

Unlike photodiode-based optoelectronic mixers, photoconductive
mixers operate through a time-varying photoconductance rather than
strongly nonlinear current transport. They therefore behave as passive
parametric frequency converters, with noise characteristics closer to
those of passive attenuating mixers than to those of actively pumped
nonlinear photodiodes.

All reported photoconductive heterodyne mixing demonstrations to date have 
been realized either with photoconductors directly integrated 
to planar antennas~\cite{DeumerAPL2025,Deumer2025}, or with photoconductors
measured directly between coplanar probes without dedicated RF/IF
circuit embedding~\cite{Peytavit2013,Tannoury2023}. In the first
case, the relevant figures of merit are primarily radiative ones
(antenna coupling efficiency, emitted or received power, free-space
bandwidth), while in the second case the photoconductor is directly
loaded by the RF and IF probes, whose out-of-band impedances remain
generally unknown. As a result, the embedding conditions seen by the
device are neither controlled nor accurately calibrated over the full
frequency range.

This partly explains why the standard RF integrated-mixer metrics such as input
return loss $S_{11}$, broadband $50\,\Omega$ impedance matching,
RF/IF port isolation, and conversion-loss optimization through
controlled embedding, have not been addressed quantitatively in
photoconductive heterodyne mixers. In particular, (i) no quantitative
on-wafer circuit-level model has been validated against measurements,
(ii) no photoconductive heterodyne mixer has been designed and
characterized as a true terahertz monolithic integrated circuit
(TMIC) with defined coplanar RF and IF ports, and (iii) the respective
contributions of the photoconductor capacitance, carrier lifetime, and
external embedding network to the conversion loss have not been
experimentally disentangled up to THz frequencies.

This work approaches photoconductive heterodyne mixing from the
perspective of RF integrated-circuit design. A fully coplanar mixer embedding is developed in
which the matching network ($R_p$), RF DC-block ($C_s$), and RF/IF
decoupling element ($C_p$) are jointly sized to provide broadband
operation, controlled RF/IF embedding conditions, and reduced RF
leakage toward the IF port. The circuit is intentionally implemented
using an unshielded planar CPW technology, which provides direct
broadband access through standard coplanar probes and simplifies
on-wafer characterization over the full frequency range.

Although such an open CPW environment does not provide the level of
electromagnetic confinement expected from mature THz MMIC/TMIC
technologies, the measured conversion losses remain in close agreement
with the hybrid distributed/lumped circuit simulations up to
\SI{500}{\giga\hertz}. This indicates that the circuit-level model captures the dominant
mixing physics and embedding effects.

The proposed architecture should therefore be regarded as a first
circuit-level validation platform rather than as a fully optimized
TMIC technology. Future implementations will require more
advanced guided structures, such as backed CPWs with ground vias or
thin-film microstrip technologies, in order to improve field
confinement and circuit reproducibility at THz frequencies. Such
evolutions remain compatible with the epitaxial layer transfer
approach used here for the fabrication of the photoconductive active
devices on high-resistivity silicon host substrates. As a proof of concept, the photoconductive layer was transferred
onto a high-resistivity SOI substrate with a $10\,\mu\mathrm{m}$
device layer, allowing the passive structures to be fabricated
directly on the SOI device layer. This configuration is compatible
with standard bulk micromachining processes, including backside
handle-wafer removal by Bosch deep reactive-ion etching (DRIE). This
opens the way to membrane-supported architectures and backside
metallization for improved THz performance.

The remainder of this paper is organized as follows.
Section~\ref{sec:model} presents the circuit-level simulation framework
and the time-varying photoconductor model. Section~\ref{sec:device}
describes the TMIC design and fabrication. Section~\ref{sec:resCirc}
presents the experimental characterization and comparison with
simulations. Section~\ref{sec:disc} discusses the implications of the
proposed approach. Finally, Section~\ref{sec:persp} investigates an
improved matching network as a possible design extension.

% =========================================================================
\section{Circuit-Level Simulation Framework}
\label{sec:model}

The proposed photoconductive heterodyne mixer is analyzed using a
hybrid circuit-level simulation framework combining small-signal
frequency-domain analysis and nonlinear time-domain integration, so
that the RF matching properties of the TMIC and the
frequency-conversion process arising from the time-varying
photoconductance can be evaluated separately within the same
topology.

The optical beat note at $f_\mathrm{LO}=|f_1-f_2|$ modulates the
carrier density in the photoconductor; for an effective carrier
lifetime $\tau$, the photoconductance
follows~\cite{Coleman1964,Peytavit2013,Peytavit2021}
\begin{equation}
 G(t)=G_0\left(
 1+
 \frac{\sin\!\left(2\pi f_\mathrm{LO}t\right)}
 {\sqrt{1+(2\pi f_\mathrm{LO}\tau)^2}}
 \right),
 \label{eq:Gt}
\end{equation}
where $G_0$ is the average photoconductance set by the total optical
power, and the instantaneous photoresistance entering the circuit
equations is $R_t(t)=1/G(t)$. The high-frequency response is governed
by two cascaded low-pass mechanisms: the intrinsic carrier-lifetime
roll-off $[1+(2\pi f\tau)^2]^{-1/2}$, and the parasitic RC roll-off
associated with the photoconductor capacitance $C_\mathrm{pc}$ loading
the local circuit impedance. The carrier-lifetime term also introduces
a phase delay, which affects only the phase of the downconverted
signal and not its amplitude, and is therefore omitted in
(\ref{eq:Gt}). The frequency conversion itself results from the
product between the RF voltage across the photoconductor and the
modulation $G(t)$, generating spectral components at the sum and
difference frequencies.

The photoconductor is embedded in a distributed RF network composed
of coplanar-waveguide (CPW) transmission lines together with lumped
matching, decoupling and biasing elements; the CPW propagation
constants are independently extracted from calibrated through-line
VNA measurements on dedicated test structures. The lengths of the distributed CPW sections are taken directly
from the mask layout, and the characteristic impedance of each
section is obtained from a quasi-TEM model \cite{Heinrich1993} of its actual
cross-section. The CPW sections adjacent to the photoconductor are
intentionally not all designed for a characteristic impedance of
$50\,\Omega$, but instead range from $40\,\Omega$ to $60\,\Omega$. This has only a minor impact on the simulated
performance but is accounted for in the model.
The section lengths and impedances are indicated in Fig.~\ref{fig:layout_circuit}.

Two circuit models are implemented in Python using the same circuit
topology and propagation parameters. The first computes the small-signal RF response in
the frequency domain using a cascaded ABCD-matrix formulation, in
which the photoconductor is represented by its average conductance
$G_0$ in parallel with the parasitic capacitance $C_\mathrm{pc}$;
this yields the scattering parameters used to evaluate impedance
matching ($|S_{11}|$) and RF/IF filtering ($|S_{21}|$). The second
engine evaluates the nonlinear frequency-conversion process with a
time-domain modified nodal analysis (MNA) solver~\cite{Ho1975} in
which the photoconductor conductance follows~(\ref{eq:Gt}). The distributed CPW sections are modeled using digital waveguide
delay lines that account for propagation delay and attenuation.
Capacitive elements are discretized using a backward-Euler companion
formulation to ensure numerical stability across the wide range of
circuit time constants. At each time step, the complete circuit
admittance matrix is rebuilt using the instantaneous value of $G(t)$,
and the transient nodal voltages are obtained by numerical
integration of Kirchhoff current equations. Simulations are
performed with a time step of $10\,\mathrm{fs}$, corresponding to
approximately 200 samples per period at $\SI{500}{\giga\hertz}$.
After removal of the initial transient regime, the IF component at
$f_\mathrm{IF}=|f_\mathrm{RF}-f_\mathrm{LO}|$ is extracted from the
simulated IF-port waveform by discrete Fourier projection, and the
conversion loss is computed as
\begin{equation}
 \mathrm{CL}=10\log_{10}
 \left(
 \frac{P_\mathrm{RF}}{P_\mathrm{IF}}
 \right),
\end{equation}
where $P_\mathrm{RF}$ is the available RF source power and
$P_\mathrm{IF}$ is the power delivered to the $50\,\Omega$ IF load.

The model is fully specified by the photoconductor parameters
$(G_0,\tau,C_\mathrm{pc})$, the matching and decoupling elements
$(R_p,C_s,C_p)$, the IF frequency, and the measured CPW propagation
constants. The nominal parameter values used throughout this work
are summarized in Table~\ref{tab:params}.
\begin{table}[t]
 \centering
 \caption{Nominal model parameters.}
 \label{tab:params}
 \footnotesize
 \setlength{\tabcolsep}{3pt}
 \renewcommand{\arraystretch}{1.05}

 \begin{tabularx}{\columnwidth}{
 @{} l
 >{\raggedright\arraybackslash}p{0.22\columnwidth}
 >{\raggedright\arraybackslash}X
 @{}
 }
 \toprule
 Symbol & Value & Meaning \\
 \midrule

 $d_\mathrm{pc}$ &
 \SI{4.0}{\micro\meter} &
 Photoconductor diameter \\

 $C_\mathrm{pc}$ &
 \SI{10.8}{\femto\farad} &
 Photoconductor capacitance \\

 $\tau$ &
 \SI{0.8}{\pico\second} &
 Effective carrier lifetime \\

 $R_\mathrm{on}=1/G_0$ &
 $240\,\Omega$ &
 Mean on-state resistance for conversion-loss characterization \\

 $R_\mathrm{on,min}$ &
 $110\,\Omega$ &
 Minimum measured on-state resistance for $S$-parameter prediction \\

 $R_p$ &
 $75\,\Omega$ &
 Ti shunt matching resistor \\

 $C_s$ &
 \SI{2}{\pico\farad} &
 RF DC-block series capacitor \\

 $C_p$ &
 \SI{1}{\pico\farad} &
 RF/IF-decoupling shunt capacitor \\

 $L_p$ &
 \SI{1}{\pico\henry} &
 Air-bridge inductance \\

 $f_\mathrm{IF}$ &
 \SI{1}{\giga\hertz} &
 Intermediate frequency \\

 \bottomrule
 \end{tabularx}
 
\end{table}

% =========================================================================
\section{TMIC Design and Fabrication}
\label{sec:device}
\begin{figure}[t]
 \centering
 \begin{subfigure}[b]{0.95\linewidth}
 \centering
 \includegraphics[width=\linewidth]{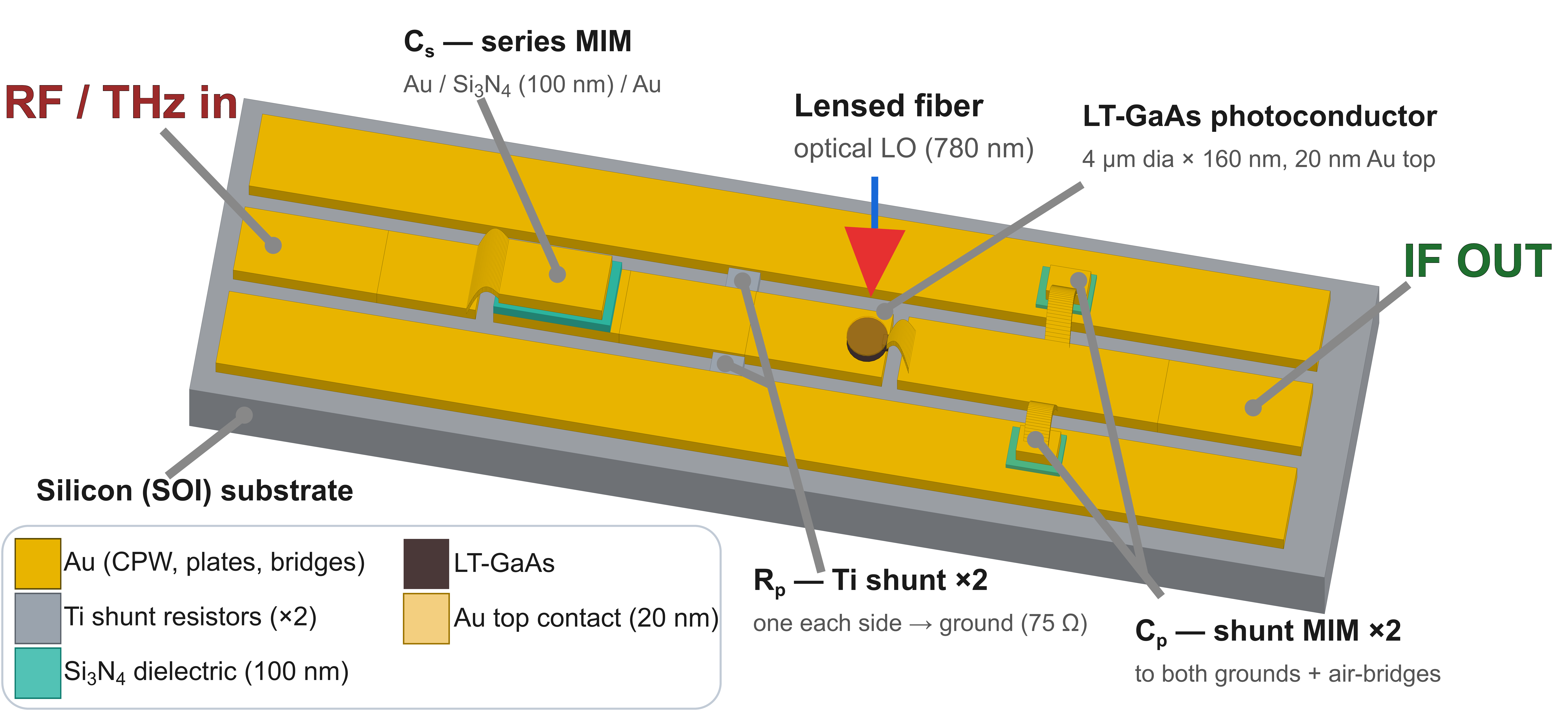}
 \caption{}
 \label{fig:layout_3d}
 \end{subfigure}

 \vspace{2mm}

 \begin{subfigure}[b]{0.95\linewidth}
 \centering
 \includegraphics[width=\linewidth]{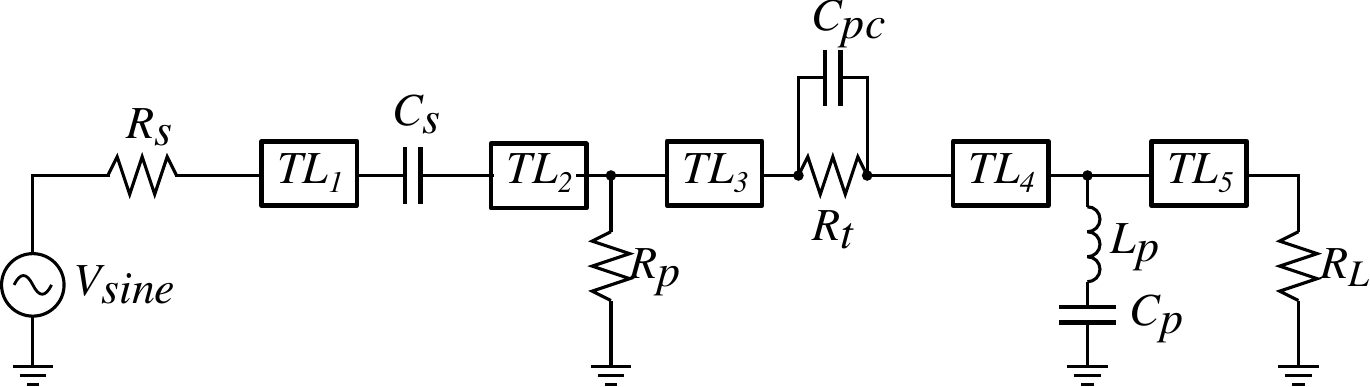}
 \caption{}
 \label{fig:layout_circuit}
 \end{subfigure}

 \caption{TMIC topology. (a)~Three-dimensional view of the mixer
 comprising the LT-GaAs photoconductor, the shunt matching resistor
 $R_p$, the series RF DC-block capacitor $C_s$, and the shunt
 RF/IF-decoupling capacitor $C_p$ embedded in a coplanar-waveguide
 environment. (b)~Equivalent circuit model combining lumped elements 
 (photoconductor, $R_p$, $C_s$, and $C_p$) with distributed
 transmission-line sections modeling the CPW accesses. The TL1--TL5
 sections have characteristic impedances of 40, 60, 60, 40, and
 40~$\Omega$, and lengths of 170, 25, 20, 84, and 191~$\mu$m,
 respectively. The air
 bridges connecting the ground planes across the shunt capacitor
 $C_p$ are modeled by the inductance $L_p$.}
 \label{fig:layout}
\end{figure}

The mixer is designed as a  TMIC with coplanar RF and IF accesses compatible with standard
on-wafer probing using $50\,\Omega$ GSG probes. The design targets
four simultaneous objectives: low input return loss, broadband
$50\,\Omega$ impedance matching, RF/IF port isolation, and minimum
conversion loss over the full operating bandwidth. Figure~\ref{fig:layout} shows the proposed circuit, in which the
photoconductor is embedded in a CPW environment together with three
lumped passive elements. A Ti thin-film shunt resistor, $R_p$, helps
match the real part of the small-signal input impedance to
$50\,\Omega$ while providing the IF termination. A series MIM
capacitor, $C_s$, blocks the DC photobias from the RF source, while a
shunt MIM capacitor, $C_p$, provides the RF return path and prevents
RF leakage into the IF circuitry. The component values were determined
using the circuit model described in Section~\ref{sec:model} and are
$R_p=75\,\Omega$, $C_s=\SI{2}{\pico\farad}$ and
$C_p=\SI{1}{\pico\farad}$; $C_s$ is sized so that the TMIC incurs no
conversion-loss penalty for $f_\mathrm{RF}\geq\SI{1}{\giga\hertz}$,
and $C_p$ so that no penalty arises as long as
$f_\mathrm{IF}\leq\SI{1}{\giga\hertz}$.

{The active device is a photomixer based on a low-temperature-grown
(LT) GaAs Fabry--Perot microcavity photoconductor, which has
demonstrated conversion losses as low as 27\,dB at 325 GHz when used as a THz
receiver~\cite{Peytavit2013} and milliwatt-level THz emission up to
325\,GHz~\cite{Peytavit2013b,Peytavit2011a}. The nominal photoconductor diameter is
\SI{4}{\micro\meter}, chosen as the best compromise between parasitic
capacitance $C_\mathrm{pc}$, ease of optical coupling, and on-state
resistance $R_\mathrm{on}$.}

\begin{figure}[t]
 \centering
 \includegraphics[width=0.95\linewidth]{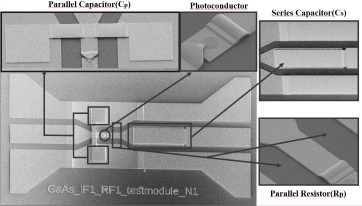}
 \caption{SEM top view of the fabricated mixer circuit.
 }
 \label{fig:fab}
\end{figure}

The circuits are fabricated on a high resistivity ($>1\,\mathrm{k}\Omega\,\mathrm{cm}$) silicon-on-insulator substrate with a \SI{10}{\micro\meter} device layer, used as a low-loss CPW
platform. The LT-GaAs active layer is grown by MBE on a GaAs wafer
with a Ga\textsubscript{0.5}In\textsubscript{0.5}P etch-stop layer,
transferred onto SOI by Au--Au thermo-compression bonding, and
released by selective wet etching of the GaAs substrate; the microcavity thickness is fine-tuned by chemical etching with intermediate reflectometry control. A \SI{20}{\nano\meter}-thick semitransparent Au layer forms the top electrode of the active device. The LT-GaAs active layer is patterned by electron-beam lithography followed by chlorine-based RIE/ICP etching, after which the underlying Au bonding layer is locally removed by Ar ion-beam etching. Ti shunt resistors, with a sheet resistance of approximately $50~\Omega/\square$, are subsequently defined by electron-beam lithography, metal evaporation, and lift-off. A first Au metallization level forms the bottom electrodes of the MIM capacitors. Si$_3$N$_4$ is then deposited over the entire wafer by PECVD, providing surface passivation, optical reflectivity control for the microcavity top electrode, and the dielectric layer of the series and shunt decoupling capacitors. Contact windows are opened in the Si$_3$N$_4$ by electron-beam lithography and dry etching. The subsequent metallization levels form the upper MIM electrodes, the CPW line sections, and the air-bridge interconnects required to connect the active device and the capacitors (Fig.~\ref{fig:fab}).

% =========================================================================

% =========================================================================
\section{Results: Full Mixer TMIC}
\label{sec:resCirc}

\subsection{$S$-parameter validation: dark measurements and operating prediction}
\label{sec:resS}
\begin{figure}[t]
 \centering
 \includegraphics[width=0.95\linewidth]{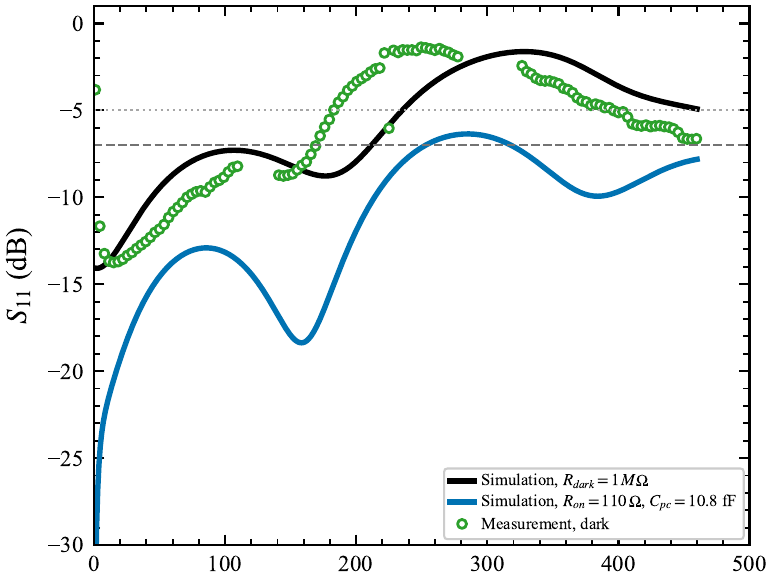}
 \caption{{Input return loss $|S_{11}|$ of the full mixer TMIC
 from \SI{0.1}{\giga\hertz} to \SI{500}{\giga\hertz}: measured in
 the dark ($R_\mathrm{off}\!\sim\!\SI{1}{\mega\Omega}$, symbols),
 simulated in the dark (solid), and simulated under operating
 conditions using the lowest measured on-state resistance
 ($R_\mathrm{on}=110\,\Omega$, dashed). The dark-state agreement
 validates the passive part of the circuit model; the operating
 curve is the corresponding in-band matching prediction.
 }}
 \label{fig:s11}
\end{figure}

\begin{figure}[t]
 \centering
 \includegraphics[width=0.95\linewidth]{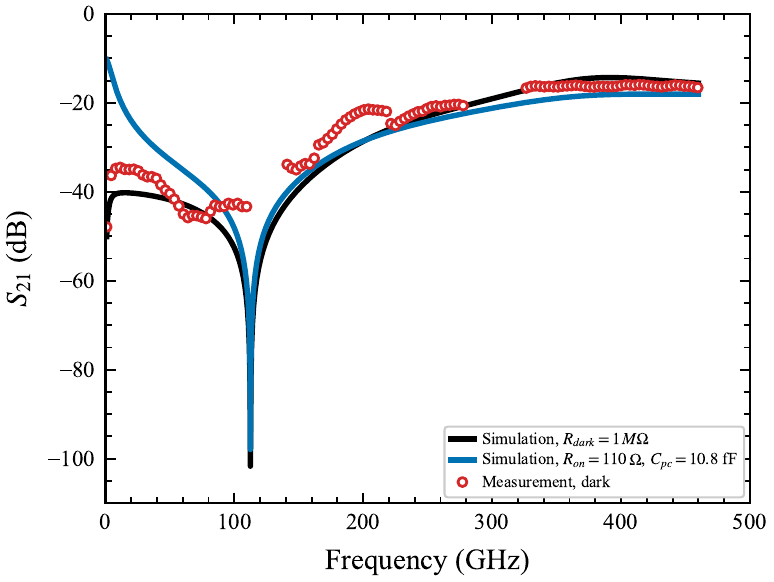}
 \caption{Measured (symbols) and simulated (solid) RF$\rightarrow$IF
 transmission $|S_{21}|$ of the full mixer TMIC in the dark, from
 \SI{0.1}{\giga\hertz} to \SI{500}{\giga\hertz}.
 }
 \label{fig:s21dark}
\end{figure}

The experimental validation is performed in two stages. First, the
passive part of the circuit model, including the CPW transmission
lines, MIM capacitors, Ti shunt resistor, and parasitic elements, is
validated by measuring the complete TMIC in the dark, where the
photoconductor behaves as an open circuit. The measurements are performed using a vector network analyzer (VNA) equipped with frequency extenders and standard GSG coplanar probes.
Calibration relies on the Line--Reflect--Reflect--Match (LRRM) method
using a commercial alumina impedance standard substrate up to
\SI{325}{\giga\hertz}, and on a multiline Thru--Reflect--Match (mTRL)
calibration performed on a commercial high-resistivity silicon
substrate above this frequency. Second, the
validated passive model is combined with the illuminated
photoconductor model to predict the circuit response under optical
excitation. Direct on-wafer $S$-parameter measurements under optical
illumination were beyond the scope of this work because they would
have required a dedicated setup combining THz VNA probing with
vertical optical excitation of the \SI{4}{\micro\meter}-diameter
photoconductor. Figures~\ref{fig:s11} and~\ref{fig:s21dark} compare the measured and
simulated $|S_{11}|$ and $|S_{21}|$ of the full TMIC over
\SIrange{0.1}{500}{\giga\hertz}, with the photoconductor modeled as
$R_\mathrm{off}\!\sim\!\SI{1}{\mega\Omega}$ in parallel with $C_\mathrm{pc}$ and the
nominal values of $C_s$ and $C_p$. For $|S_{21}|$, a series
inductance $L_p=\SI{1}{\pico\henry}$, representing the air bridges
crossing the shunt capacitor $C_p$, was added to the model to
further improve the agreement; this value is fully consistent with
the bridge dimensions (\SI{12}{\micro\meter} long,
\SI{10}{\micro\meter} wide, \SI{0.6}{\micro\meter} thick). Across
the entire span, the amplitude agreement is within a few dB and
reproduces the main features of the measurements (resonant dips,
broad envelope), confirming that the passive embedding network is
correctly captured by the model up to \SI{500}{\giga\hertz}.

For $|S_{11}|$, however, a frequency offset is observed between
the measured and simulated peaks and troughs. This is tentatively
attributed to parasitic capacitive contributions in the transitions
between the different circuit elements, which are not accounted for in
the model, since the circuit is represented using only lumped elements
and CPW line sections. Such parasitic elements, which are not included in the circuit model,
shift the resonant frequencies while leaving the overall amplitude
response essentially unchanged.

{The same model is then exercised under operating conditions to
predict the input return loss under optical bias, and the resulting
curve is superimposed on the dark-state data in Fig.~\ref{fig:s11}.
The on-state resistance used for this prediction is
$R_\mathrm{on}=110\,\Omega$, the lowest value measured on the
fabricated devices, obtained at the maximum optical pump power
applied without observable short-term degradation. Since the long-term stability of the device under high optical pump power was not investigated, the minimum measured $R_\mathrm{on}$ is taken as the best-case operating point for the matching analysis, whereas the conversion-loss characterization of Section ~\ref{sec:resCL} is performed at reduced optical power
($R_\mathrm{on}\approx240\,\Omega$). Under these conditions
$|S_{11}|$ stays below $-6.5$\,dB on the whole frequency band. While this is short of an
ideal $|S_{11}|<-10$\,dB target across the full band, it remains a
usable matching performance for a THz mixer operating over more
than a decade of bandwidth. \rev{A resistive series--shunt evolution of
the matching network, trading conversion loss for return-loss
flatness, is analyzed in Section~\ref{sec:persp}.}}

\subsection[Conversion loss measurement setup]{\rev{Conversion loss measurement setup}}
\label{sec:meas}
\begin{figure*}[t]
 \centering
 \includegraphics[width=0.98\linewidth]{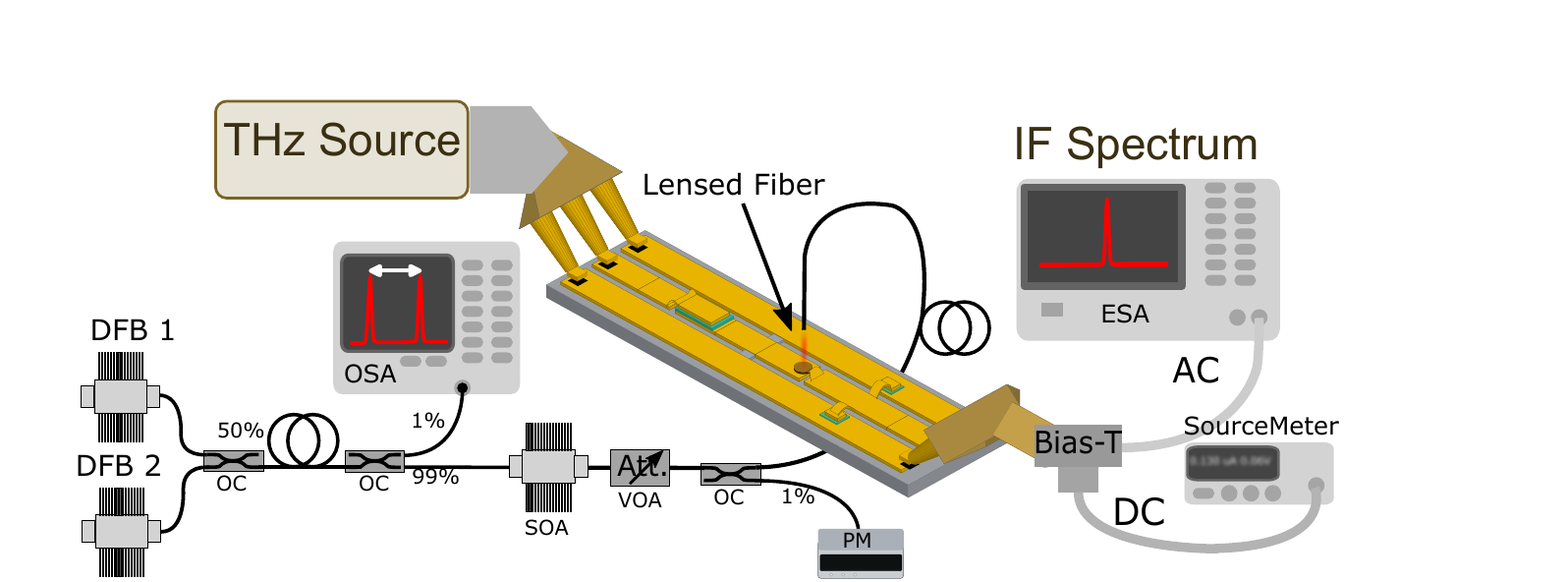}
 \caption{On-wafer Conversion Loss measurement setup.\textbf{DFB}: distributed-feedback laser; \textbf{OC}: optical coupler; \textbf{OSA}: optical spectrum analyzer; \textbf{SOA}: tapered semiconductor optical amplifier; \textbf{VOA}: variable optical attenuator; \textbf{PM}: optical power meter; \textbf{ESA}: electrical spectrum analyzer; \textbf{Bias-T}: bias tee.}
 \label{fig:setup}
\end{figure*}

The optical local oscillator (LO) is formed by the optical beat note generated from two temperature-tunable fiber-coupled \SI{780}{\nano\meter} distributed-feedback (DFB) lasers. The two optical carriers are combined using a 50:50 fiber coupler, while an integrated temperature-control system independently tunes each laser wavelength to set the beat frequency.The combined optical signal is subsequently amplified by a tapered semiconductor optical amplifier (SOA). Finally, the optical LO is delivered to the photoconductor through a lensed fiber with a \SI{4}{\micro\meter} mode-field diameter (MFD), focused normally onto the microcavity. The RF/THz signal is provided
by an electronic source and VDI multipliers covering $0$--\SI{500}{\giga\hertz}
in successive bands, fed to the RF port by waveguide-coupled GSG coplanar probes. The
IF at \SI{1}{\giga\hertz} is collected by a coaxial-coupled GSG coplanar probe and read
on a spectrum analyzer. For each RF frequency the optical beat note is
tuned to $f_\mathrm{LO} = f_\mathrm{RF} - \SI{1}{\giga\hertz}$.
\rev{The overall measurement setup, combining the fiber-delivered
optical LO, the multiplied RF/THz source and the coplanar RF and IF
probing, is shown in Fig.~\ref{fig:setup}.}

The available RF power is measured in the waveguide with a VDI
Erickson PM5 power meter and referred to the circuit input by
subtracting the RF-probe insertion loss ($S_{21}$, from the
manufacturer data), which depends on the frequency band. On the IF
side, the insertion losses of the IF probe and of the cables between
the circuit output and the spectrum-analyzer input are subtracted so
that the reported IF power corresponds to the power at the circuit
output. This de-embedding is treated as scalar, i.e.\ only the
$S_{21}$ magnitude is accounted for, under the assumption of ideal
matching at the probe tips, which is justified by the low reflection
coefficient specified for the probes ($|S_{11}|,|S_{22}| 
<-15$\,dB). The reported figures thus correspond to the available RF
power at the circuit input and to the IF power at the circuit
output.

{Conversion-loss characterization is performed at moderate
optical pump power ($\approx\SI{40}{\milli\watt}$), corresponding to an on-state photoresistance $R_\mathrm{on}\approx240\,\Omega$, to keep the device safely
well below its degradation threshold throughout the measurement campaign; this
value is used in conversion-loss simulations. Higher optical pump powers, up to approximately \SI{95}{\milli\watt} (corresponding to $R_\mathrm{on}\approx\SI{110}{\ohm}$), were applied without observable short-term degradation. This minimum measured value was retained for the $S$-parameter predictions of Section~\ref{sec:resCirc}.}

\subsection[Conversion loss results]{\rev{Conversion loss results}}
\label{sec:resCL}
Figure~\ref{fig:cl_tau} shows the measured conversion loss of the full
mixer TMIC up to 500 GHz, which constitutes, to our knowledge,
the first quantitative on-wafer characterization of an optoelectronic
mixer circuit over such a broad band. It is
\SIrange{22.5}{25.5}{\decibel} below \SI{320}{\giga\hertz} and increases
to \SIrange{26}{34}{\decibel} up to \SI{500}{\giga\hertz}, with a
typical variability induced by IF/RF frequency sweep of $\pm\SI{1}{\decibel}$.
 
The high-frequency roll-off is governed by two physical parameters:
 the capacitance of the photoconductor $C_\mathrm{pc}$ and the carrier
lifetime $\tau$. {Instead of treating both as free fit
parameters, the capacitance is fixed to the calculated value from the
device geometry and only the lifetime is adjusted, keeping the
extraction physically constrained.}
 
{The photoconductor capacitance is computed from the actual
top-electrode area extracted from the layout,
\SI{13.9}{\micro\meter\squared} (a nominally
\SI{4}{\micro\meter}-diameter disk merged with its contact pad), over
a \SI{162}{\nano\meter}-thick GaAs cavity with $\varepsilon_r=12.9$.
Adding the edge fringing-field contribution following the closed-form
correction for a circular plate~\cite{Sloggett1986}, the estimated
capacitance is $C_\mathrm{pc}\approx\SI{10.8}{\femto\farad}$.}
 
Time-resolved photoreflectance measurements performed by the authors on LT-GaAs epitaxial layers grown under equivalent conditions yield carrier lifetimes of approximately $\tau\approx\SI{0.5}{\pico\second}$. This value is used as the lower bound of the investigated range, since a longer effective response time is expected under CW optical pumping induced by a non-negligible trap emptying time.
{Figure~\ref{fig:cl_tau} compares the measurement with simulations
for $\tau$ swept between \SI{0.5}{\pico\second} and
\SI{1.0}{\pico\second} at the computed capacitance
\rev{$C_\mathrm{pc}=\SI{10.8}{\femto\farad}$}. The best agreement is obtained
for $\tau=\SI{0.8}{\pico\second}$. Figure~\ref{fig:cl_cpc} then fixes the
lifetime at this value and varies $C_\mathrm{pc}$ between
\SI{10}{\femto\farad} and \SI{12}{\femto\farad}, yielding the best agreement for {$C_\mathrm{pc}=\SI{11.25}{\femto\farad}$} close to the calculated value. This second sweep is intended  to check consistency rather than to provide a rigorous extraction, which would require a joint two-dimensional fit in the $(\tau, C_\mathrm{pc})$ plane.  As can be observed from the relatively small separation between the simulated curves in Fig.~\ref{fig:cl_cpc}, which remains smaller than both the intra-band experimental variations and the measured ripples, the influence of $C_\mathrm{pc}$ on the simulated conversion loss is relatively weak compared with the overall experimental uncertainty. These uncertainties arise primarily from the absence of a complete VNA characterization of the THz signal path between the source and the on-wafer probe. As a result, the modeling assumes an ideal $50\,\Omega$ source and perfectly matched probes, accounting only for their measured insertion loss ($S_{21}$). In practice, however, the impedance presented by the probes to the circuit is expected to vary with frequency and is unlikely to remain purely real over the entire frequency range. Consequently, extracting $C_\mathrm{pc}$ with significantly higher accuracy would not be justified by the present experimental conditions.
 
\begin{figure}[t]
 \centering
 \includegraphics[width=0.95\linewidth]{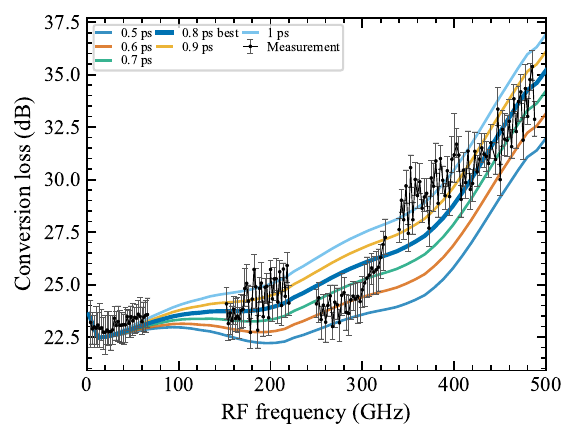}
 \caption{{Measured conversion loss (symbols) versus simulation
 for $\tau$ swept between \SI{0.5}{\pico\second} and
 \SI{1.0}{\pico\second}, at the computed capacitance
 $C_\mathrm{pc}=\SI{10.8}{\femto\farad}$.}}
 \label{fig:cl_tau}
\end{figure}
 
\begin{figure}[t]
 \centering
 \includegraphics[width=0.95\linewidth]{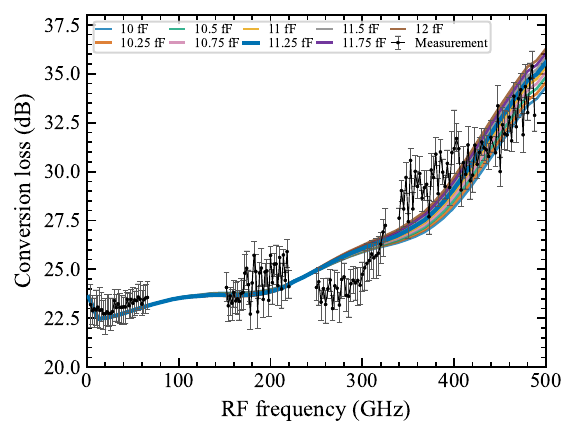}
 \caption{{Measured conversion loss (symbols) versus simulation
 for $C_\mathrm{pc}$ swept between \SI{10}{\femto\farad} and
 \SI{12}{\femto\farad}, at fixed best-fit lifetime
 $\tau=\SI{0.8}{\pico\second}$.}
 }
 \label{fig:cl_cpc}
\end{figure}

% =========================================================================

\section{Discussion}
\label{sec:disc}
Table~\ref{tab:soa} summarizes representative heterodyne mixers in
the THz range, including commercial Schottky modules and academic
photoconductive and UTC-PD demonstrations, in terms of frequency
coverage, conversion loss and local-oscillator technology. Electronic
Schottky mixers remain unmatched in absolute conversion loss
(\SIrange{8}{14}{\decibel} for fundamental and sub-harmonic modules)
but operate over waveguide-limited bands and require a
high-frequency electronic LO chain, while even-harmonic mixers relax
the LO constraint at the price of a conversion loss comparable to or
higher than that of photoconductive mixers. Previously reported
photoconductive heterodyne demonstrations achieve
\SIrange{20}{40}{\decibel} of conversion loss over very wide bands
with an optically delivered LO; they were characterized either in
free-space through an integrated antenna, or on-wafer with the bare
photoconductor contacted directly between probes, without a defined
RF/IF circuit embedding. The conversion loss of the present TMIC is
comparable to the best of these demonstrations, with the difference
that it is obtained in a fully characterized $50\,\Omega$
coplanar environment, validated by a circuit model up to
\SI{500}{\giga\hertz}. This is precisely the configuration required
when the mixer is to be inserted into a larger system, whether an
on-wafer test instrument or a packaged receiver front-end.

\begin{table*}[t]
 \centering
 \caption{Representative state-of-the-art heterodyne mixers in the
THz range, including academic demonstrations and commercially
available Schottky modules. SHM: sub-harmonic mixer; FM: fundamental
mixer; EHM: even-harmonic mixer; PC: photoconductive; UTC-PD: uni-traveling-carrier photodiode;
WG: waveguide; CL: conversion loss; iso.: RF/IF isolation; n/r: not
reported. Commercial mixer specifications from VDI are generally
reported in double-sideband (DSB) conditions; the values reported here
include a +3\,dB correction for comparison with the present
single-tone conversion-loss definition.}
 \label{tab:soa}
 \footnotesize
 \setlength{\tabcolsep}{4pt}
 \begin{tabular}{@{}l l c c >{\centering\arraybackslash}p{2.2cm} l l@{}}
 \toprule
 Reference & Type & Band & CL\,[dB] & $|S_{11}|$ / iso. & Coupling & LO \\
 \midrule
 \cite{Thomas2005} & Schottky SHM & 300--360\,GHz & 8--10 & n/r  & WG module & electronic \\
 \cite{VDI_mixers} VDI comm.\ SHM & Schottky SHM & 220--500\,GHz & 11--13 & n/r  & WG module & electronic \\
 \cite{VDI_mixers} VDI comm.\ FM & Schottky FM & 220--500\,GHz & 13--14 & n/r  & WG module & electronic \\
 \cite{VDI_mixers} VDI comm.\ EHM & Schottky EHM & 220--500\,GHz & 30--45 & n/r  & WG module & low-freq.\ elec. \\
 \cite{Peytavit2013} & Photoconductive mixer & $\sim$100\,GHz & 20 & n/r &     Uncontrolled probe embedding& optical \\
 \cite{Tannoury2023} & Plasmonic-cavity PC & 0.1--0.5\,THz & 20--35 & n/r &     Uncontrolled probe embedding& optical \\
 \cite{DeumerAPL2025} & Waveguide-coupled PC & 0.1--0.6\,THz & 20--40 & n/r & free-space & optical \\
 \cite{Rouvalis2012} & UTC-PD & 30--110\,GHz & 20--30 & n/r &     Uncontrolled probe embedding & optical (+shot noise) \\
 \cite{BelioApaolaza2022} & Schottky + opt.\ LO & 300\,GHz & 14 & n/r & WG module & optical \\
 \textbf{This work} & \textbf{PC mixer TMIC} & 0.001--0.5\,THz & 20--34 & \textbf{sim.\ $R_\mathrm{on}$ / meas.\ $R_\mathrm{off}$} & \textbf{coplanar TMIC} & optical \\
 \bottomrule
 \end{tabular}
\end{table*}

The coplanar access adopted in this work is intended primarily for
on-wafer characterization. The SOI technology also provides a
straightforward route toward packaged implementations. By locally
removing the silicon handle wafer beneath the active circuit, the
\SI{10}{\micro\meter} device layer can be used as a suspended
dielectric platform, enabling alternative coupling schemes beyond the
coplanar access adopted here.
Alternatively, the same membrane process supports heterogeneous
integration on micromachined silicon coplanar test probes, in which
the TMIC, retaining its $G(t)$-based parametric mixing function, is
reported on a silicon probe tip to provide a broadband on-wafer
spectrum analyzer with the optical LO delivered by fiber. While
these transpositions are not demonstrated here and will require
dedicated electromagnetic redesign of the input-coupling structures,
the SOI-on-membrane platform is consistent with both, and the TMIC
topology and parametric model validated above carry over without
modification.

% =========================================================================
% NEW SECTION (revision) --- wrapped in {...}
% =========================================================================
\section[Perspective: Broadband --10 dB Matching with a Series--Shunt Resistive Network]{{Perspective: Broadband $-10$\,dB Matching with a Series--Shunt Resistive Network}}
\label{sec:persp}

{\rev{The simulated return loss of the fabricated design under
operating conditions (Fig.~\ref{fig:s11}) exhibits a residual
mismatch at the upper end of the band, driven by the capacitive
loading of $C_\mathrm{pc}$ at the photoconductor node.} A purely
shunt-resistive matching network ($R_p$ alone) cannot bound the input
reflection coefficient once the node impedance departs from its
low-frequency value. A simple circuit-level evolution, fully
compatible with the present technology, consists in complementing the
shunt resistor with a thin-film \emph{series} matching resistor
$R_\mathrm{s,m}$ inserted between the RF access and the
photoconductor node.}

The principle is that of classical dissipative (resistive)
padding: the series resistor partially isolates the RF port from the
frequency-dependent impedance of the loaded photoconductor node, so
that the reflection coefficient remains bounded even where the node
impedance collapses under the loading of $C_\mathrm{pc}$ or rotates
because of the embedding parasitics. The shunt resistor is
simultaneously increased so that the low-frequency input resistance,
now $R_\mathrm{s,m}$ in series with
$R_p\,\|\,R_\mathrm{on}$, remains close to the $50\,\Omega$
target. Using the circuit model validated in
Section~\ref{sec:resCirc}, the retained values are
$R_\mathrm{s,m}=20\,\Omega$ and $R_p=100\,\Omega$ (instead of
$R_p=75\,\Omega$ without series resistor), all other elements
being kept identical to the fabricated design.

To quantify the cost of matching, Figs.~\ref{fig:penalty_nom}
and~\ref{fig:penalty_opt} report, for the fabricated nominal network
and for the proposed series--shunt network, respectively, the
conversion-loss penalty $\Delta\mathrm{CL}$ with respect to the
unmatched case, corresponding to the bare photoconductor embedding
($R_p$ removed and no series resistor), together with the simulated
RF-port return loss in both cases. The unmatched device naturally
exhibits the lowest conversion loss because the matching network
considered here is purely dissipative. However, its return loss
remains poorly controlled, varying between approximately $-4$ and
$-9$\,dB across the band.

{Even the fabricated design already embodies this trade-off: the
nominal shunt resistor $R_p=75\,\Omega$ costs up to
$\approx\SI{2.8}{\decibel}$ of conversion loss at low frequency,
$\approx\SI{1.5}{\decibel}$ around
\SIrange{100}{200}{\giga\hertz}, and less than \SI{1}{\decibel} above
\SI{300}{\giga\hertz}, while keeping $|S_{11}|$ below $-10$\,dB only
up to about \SI{215}{\giga\hertz}
(Fig.~\ref{fig:penalty_nom}). With
$R_\mathrm{s,m}=20\,\Omega$ and $R_p=100\,\Omega$
(Fig.~\ref{fig:penalty_opt}), the simulated return loss remains below $-10$\,dB over the entire \SIrange{0.001}{0.5}{\tera\hertz} band (worst case
$\approx-9.5$\,dB near \SI{90}{\giga\hertz} and
\SI{290}{\giga\hertz}), suppressing in particular the poorly matched
region of the nominal design in the upper part of the band. The
price is a conversion-loss penalty of
\SIrange{2.8}{4.5}{\decibel} across the band relative to the
unmatched device, i.e.\ \SIrange{1}{3}{\decibel} more than the
nominal design over most of the upper half of the band.}

With a photoconductor whose on-state resistance and parasitic
capacitance are fixed by the optical pump power and device geometry,
broadband resistive matching introduces an unavoidable conversion-loss
penalty. The preferred trade-off is therefore application dependent. For measurement front-ends such as the on-wafer
spectrum-analyzer probe discussed in
Section~\ref{sec:disc}, a flat and well-controlled input match
suppresses standing waves along the access structures, thereby
improving measurement accuracy directly. The associated increase in
conversion loss remains smooth and limited to a few decibels, making
it straightforward to calibrate out. Under these conditions, the
series--shunt network is clearly preferable. By contrast, for
sensitivity-critical receivers, the nominal network, or even the
unmatched device combined with a reactive band-limited matching
network, remains the preferred solution.

{From a fabrication standpoint, the series resistor requires no
additional process step: it can be implemented in the same
$\sim50\,\Omega/\square$ Ti thin film already used for the shunt
resistor $R_p$, by inserting a short resistive section in the CPW
center conductor on the RF side. This series--shunt resistive
matching network is therefore the natural evolution of the present
topology for the next fabrication run.}

\begin{figure}[t]
 \centering
 \includegraphics[width=0.95\linewidth]{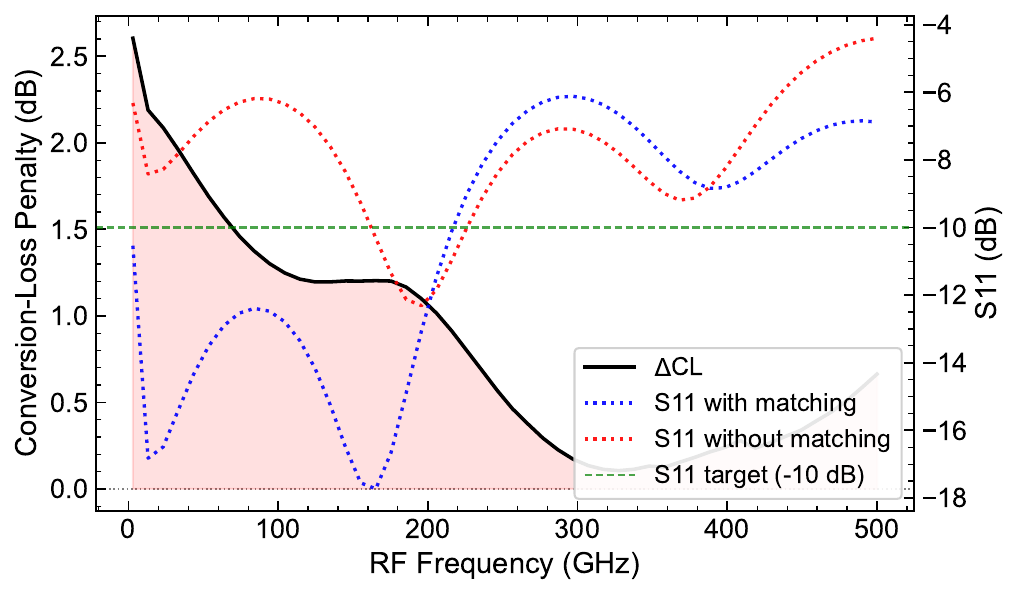}
 \caption{\rev{Matching penalty of the fabricated nominal network
 ($R_p=75\,\Omega$, no series resistor): conversion-loss
 penalty $\Delta\mathrm{CL}$ with respect to the unmatched case
 (solid, left axis) and simulated RF-port return loss with and
 without the matching network (dotted, right axis); the
 $-10$\,dB target is indicated (dashed horizontal line).}}
 \label{fig:penalty_nom}
\end{figure}

\begin{figure}[t]
 \centering
 \includegraphics[width=0.95\linewidth]{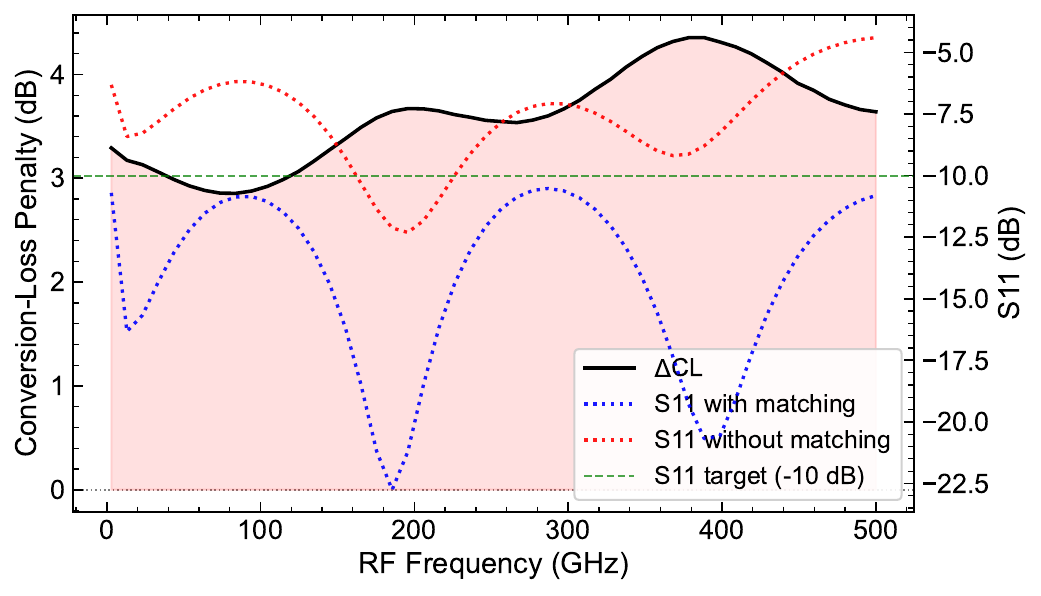}
 \caption{\rev{Matching penalty of the proposed series--shunt
 dissipative matching network
 ($R_\mathrm{s,m}=20\,\Omega$, $R_p=100\,\Omega$):
 conversion-loss penalty $\Delta\mathrm{CL}$ with respect to the
 unmatched case (solid, left axis) and simulated RF-port return
 loss with and without the matching network (dotted, right axis);
 the return loss remains close to or below the $-10$\,dB target
 (dashed horizontal line) over the full band.}}
 \label{fig:penalty_opt}
\end{figure}

% =========================================================================
\section{Conclusion}
\label{sec:concl}
A coplanar THz optoelectronic mixer has been designed,
fabricated, and characterized on-wafer as a terahertz monolithic
integrated circuit (TMIC). Unlike previous photoconductive heterodyne
demonstrations based on antenna coupling or direct probe measurements,
the present work treats the mixer as an integrated RF circuit with
controlled input matching, RF/IF isolation, and conversion-loss
optimization. The passive part of the circuit
model is validated against $|S_{11}|$ and $|S_{21}|$ VNA measurements
of the unilluminated TMIC ($R_\mathrm{off}\!\sim\!\SI{1}{\mega\Omega}$)
up to \SI{500}{\giga\hertz}; the same model under operating
conditions ({$R_\mathrm{on}\!\sim\!110\,\Omega$, the lowest
on-state resistance measured at the maximum applied optical pump
power}) predicts a
simulated $|S_{11}|$ below $-10$\,dB up to \SI{215}{\giga\hertz} and
a maximum of about $-6.5$\,dB near \SI{290}{\giga\hertz}.

A parametric circuit model predicts the conversion loss of the full TMIC within
$\pm \SI{1}{\decibel}$ over the entire \SIrange{1}{500}{\giga\hertz}
band, with \rev{$\tau \approx \SI{0.8}{\pico\second}$
 and the geometry-computed
$C_\mathrm{pc} \approx \SI{10.8}{\femto\farad}$}. The
conversion loss is \SIrange{22.5}{25.5}{\decibel} below
\SI{320}{\giga\hertz} and \SIrange{26}{34}{\decibel} up to
\SI{500}{\giga\hertz}. The SOI platform employed in this work is also compatible with future
packaged implementations and integration with micromachined silicon RF
technologies. {Circuit simulations
based on the validated model further show that a series--shunt
resistive matching network ($R_\mathrm{s,m}=20\,\Omega$,
$R_p=100\,\Omega$) would maintain $|S_{11}|$ close to or below
$-10$\,dB over the entire band, at the cost of a
\SIrange{2.8}{4.5}{\decibel} conversion-loss penalty relative to the
unmatched photoconductor --- an explicit, application-dependent
trade-off between return-loss flatness and sensitivity that provides
a straightforward improvement path for the next design iteration.}
Together, these results demonstrate that photoconductive heterodyne
mixers can be designed, analyzed, and optimized using established
microwave circuit techniques, providing a foundation for future
integrated THz optoelectronic receivers.

% =========================================================================
\section*{Acknowledgment}
The authors acknowledge the support of the French Ministry of Higher
Education and Research, the French National Research Agency (ANR), the
Hauts-de-France Regional Council, the University of Lille, the
Initiative of Excellence of the University of Lille, the European
Metropolis of Lille (MEL), the French National Centre for Scientific
Research (CNRS), and the European Regional Development Fund (ERDF).
The authors thank the IEMN cleanroom staff for fabrication support, and Jean-François Lampin for fruitful discussions.

% =========================================================================
\bibliographystyle{IEEEtran}
\bibliography{references}

\end{document}